\documentclass[fontsize=11pt, paper=letter, twocolumn]{scrartcl}
\usepackage[utf8]{inputenc}
\usepackage[english]{babel}
\usepackage{authblk}
\usepackage{siunitx}
\usepackage{nicefrac}
\usepackage{amsmath}
\usepackage{cuted}
\usepackage{graphicx}
\usepackage[includeheadfoot, headsep=.1in, margin={.8in,.6in}]{geometry}
\usepackage{hyperref}

\newcommand{\vc}[1]{\ensuremath{\mathbf{#1}}} 

\title{Single-cell diffraction tomography\\with optofluidic rotation about a tilted axis} 

\author{Paul M\"uller\footnote{To whom correspondence should be addressed.}, Mirjam Sch\"urmann, Chii J. Chan, and Jochen Guck}

\affil{Biotechnology Center of the TU Dresden, Germany}

\date{September 8, 2015}

\begin{document} 

\maketitle 

\begin{strip}
This work has been presented at SPIE Nanoscience + Engineering (Proceedings Volume 9548, Optical Trapping and Optical Micromanipulation XII, 95480U, 2015) and has been published at \url{https://dx.doi.org/10.1117/12.2191501}.

© 2015 Society of Photo Optical Instrumentation Engineers (SPIE). One print or electronic copy may be made for personal use only. Systematic reproduction and distribution, duplication of any material in this publication for a fee or for commercial purposes, or modification of the contents of the publication are prohibited. 
\end{strip}

\abstract{
Optical diffraction tomography (ODT) is a tomographic technique that can be used to measure the three-dimensional (3D) refractive index distribution within living cells without the requirement of any marker. In principle, ODT can be regarded as a generalization of optical projection tomography which is equivalent to computerized tomography~(CT). Both optical tomographic techniques require projection-phase images of cells measured at multiple angles. However, the reconstruction of the 3D refractive index distribution post-measurement differs for the two techniques.
It is known that ODT yields better results than projection tomography, because it takes into account diffraction of the imaging light due to the refractive index structure of the sample.
Here, we apply ODT to biological cells in a microfluidic chip which combines optical trapping and microfluidic flow to achieve an optofluidic single-cell rotation. In particular, we address the problem that arises when the trapped cell is not rotating about an axis perpendicular to the imaging plane, but instead about an arbitrarily tilted axis. In this paper we show that the 3D reconstruction can be improved by taking into account such a tilted rotational axis in the reconstruction process.
}

\section{INTRODUCTION}
\label{sec:intro}
Techniques that measure the refractive index of biological cells always require quantitative phase measurements. The phase shift introduced to the imaging beam by biological cells is typically in the order of~$\nicefrac{\pi}{2}$. The cause for the phase shift is the mostly real-valued refractive index of cells, which is complex-valued in general.
Examples for quantitative phase imaging techniques are refractive index matching or interference-based digital holographic microscopy (DHM) \cite{Barer1957, Schuermann2015143}. In order to achieve a 3D reconstruction of the refractive index within biological cells, a tomographic approach is required. 

Refractive index reconstruction with a tomographic approach can be addressed with optical projection tomography \cite{Sharpe2004, Charriere2006}.
The difference between a tomographic approach and conventional 3D imaging techniques, such as confocal imaging, lies in the way the data is acquired and processed. In confocal microscopy, the 3D volume is scanned by a laser beam, whereas in tomography 2D projections of the sample are recorded from different angles. Therefore, tomographic approaches require more elaborate post-processing steps to reconstruct the 3D volume from the 2D projections.
An efficient technique to reconstruct the refractive index from phase projections is the backprojection algorithm \cite{Kak2001, Devaney1982, Mueller15a}. However, the backprojection algorithm is only applicable for imaging wavelengths that are small (e.g. x-ray radiation) compared to the sample. When imaging biological samples with wavelengths of e.g. 400$\,$nm or above, diffraction occurs and the reconstruction by backpojection becomes blurry\cite{Mueller15}.
The theory of diffraction tomography addresses the wave nature of light with the Rytov approximation\cite{Kak2001, Mueller15a}. As a result, the above mentioned blurring artifacts do not appear when the corresponding \textit{backpropagation} algorithm is used to reconstruct the refractive index.

Here, we combine diffraction tomography with a microfluidic setup that uses optical forces to trap, and background flow to rotate single cells in suspension\cite{Kolb_2014}. In practice, cells do not always rotate about an axis perpendicular to the imaging axis due to asymmetry in their shape or slight misalignments of the setup. We address this issue computationally and show that it is possible to improve the image contrast of the reconstructed refractive index distribution.

\section{Methods}
\subsection{Optofluidic cell rotation}
\begin{figure*}
\includegraphics[width=\textwidth]{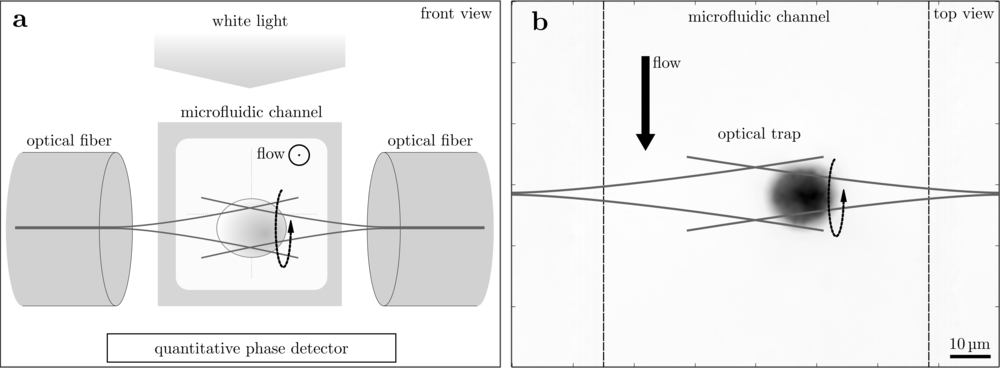}
\caption
{ \label{fig:1} 
\textbf{a)} The sketch illustrates the working principle of optofluidic cell rotation. The cell is trapped in a dual beam laser trap created by two opposing optical fibers, which results in a fixed position of the cell in the microfluidic channel (cross-section shown). When flow is introduced (arrow pointing out of the plane of this figure), the cell starts to rotate, because the flow in the center of the channel is high compared to the flow at the walls. During rotation, the cell is imaged with white light. Amplitude and phase are recorded with a quantitative phase imaging device (see text). \textbf{b)} The top view shows the measured phase image overlaid with a schematic drawing of the optical trap illustrating the rotation resulting from the flow in the microfluidic channel.
}
\end{figure*}

We achieve single-cell rotation by combining a microfluidic channel with a dual beam laser trap, as presented by Kolb et al.\cite{Kolb_2014}. The working principle is illustrated in figure~\ref{fig:1}a.
To achieve the dual beam laser trap, two opposing optical fibers are aligned along a single axis. In between the two fibers, the microfluidic channel, a square shaped glass capillary, is introduced. The center of the dual beam laser trap is located in the bottom half of the channel.
This positioning has two reasons: first, cells that drift to the bottom of the channel due to gravity can be successfully trapped and second, with flow in the channel, a trapped cell experiences an asymmetric flow velocity field, causing it to rotate. Figure~\ref{fig:1}b shows the phase image of a trapped HL60/S4 myeloid precursor cell in the microfluidic channel.

\subsection{Image acquisition}
As discussed in the introduction, quantitative phase images are required for refractive index tomography. Image acquisition is performed perpendicular to both microfluidic channel and optical trap as shown in figure~\ref{fig:1}a. Here, we use a quantitative phase imaging camera `SID4BIO' from Phasics S.A. (Saint Aubin, France) to measure the amplitude $A$ and the phase $\Phi$ changes of white light that passes through the cells. The camera employs quadriwave shearing interferometry \cite{Chanteloup_2005} to measure the wave field $u(\vc{r})=A(\vc{r})\exp(i\Phi(\vc{r}))$. Before reconstructing the refractive index, the measured images must be background corrected and aligned. Background correction is performed with linear ramp filters in the phase images and background subtraction in the amplitude images. Alignment with respect to the center of a rotating cell is achieved by fitting a circle to the contour of the cell in the phase images\cite{Schuermann2015143}.

\subsection{Determination of tilted axis}
\begin{figure*}
\includegraphics[width=\linewidth]{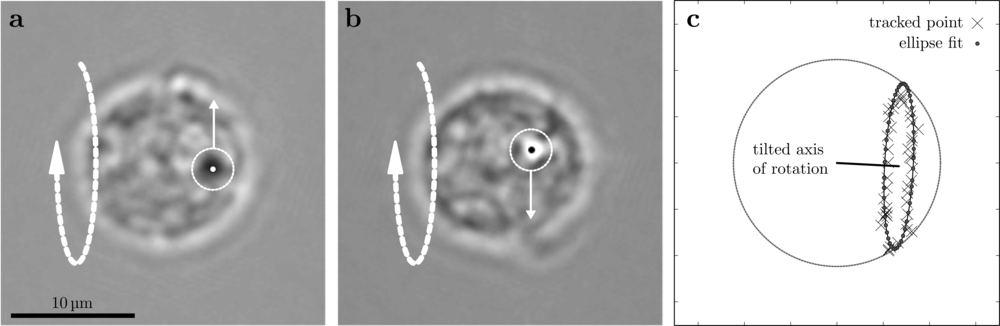}
\caption
{ \label{fig:2} 
\textbf{a,b)} The intensity images are shown for two different rotational positions of an HL60/S4 myeloid precursor cell. The diffraction spot resulting from a strongly diffracting part of the cell, black in (a) and white in (b), is required for tracking the rotational position of the cell. \textbf{c)} Those frames that allowed tracking of this diffraction spot contribute points to the 2D fit of an ellipse. With this fit and the knowledge of the axial position of the feature creating the diffraction spot, it is possible to estimate the 3D~orientation of the cell in all frames (see text). The white point in (a) and the black point in (b), located in the highlighted circles, correspond to the points on the ellipse after fitting.
}
\end{figure*}

In order to determine the orientation of the rotational axis \vc{r_A} from the acquired wave fields, we apply a particle-tracking approach. First, a part of the cell which has a high refractive index and creates a diffraction spot in the amplitude image $A(\vc{r})$ is registered and tracked with a particle tracking algorithm\cite{DanielB_2014}. Figures~\ref{fig:2}a and~\ref{fig:2}b show the cell for two rotational positions with the diffraction spot following the rotation of the cell along an ellipse. Figure~\ref{fig:2}c illustrates the path of the diffraction spot for a full rotation from 0 to 360$^\circ$ with a total number of 51~frames. If the rotational axis was not tilted with respect to the imaging plane, then all tracked spots would reside on a line.
However, the tracked spots are distributed along an ellipse which we determine by a least-squares fit. The images shown in figure~\ref{fig:2} are aligned with respect to the center of the cell, as discussed above.
Thus, we assume that the ellipse is a projection of a circle on the surface of a centered sphere onto the 2D detector plane. 
From this elliptical fit, we obtain the tilt angle in the imaging plane and the magnitude of the tilt angle perpendicular to the imaging plane. The missing information about the sign of the tilt angle perpendicular to the imaging plane is obtained with numerical focusing\cite{nrefocus}. Numerical focusing is a well-known technique that simulates the propagation of a complex wave $u(\vc{r})$ in free space, resulting in images comparable to those obtained by tuning the focus of a microscope.
For each tracked point in the 2D ellipse (figure~\ref{fig:2}c), we observed that the axial position of the diffraction spot highlighted in figure \ref{fig:2}a and \ref{fig:2}b is white when it points at the observer (focused forward) and black when it points away from the observer (focused backward). 
We use this change in intensity to determine the sign of the tilt of the rotational axis \vc{r_A} introduced above.
Additionally, we overcome the problem of unknown rotational cell positions for subsets of frames that do not allow tracking of the diffraction spot by assigning equally-spaced rotational positions.

\subsection{Refractive index reconstruction}
The theoretical foundation of diffraction tomography based on the Rytov approximation and the implementation of the backpropagation algorithm that we describe elsewhere\cite{Mueller15a, Mueller15}, discusses the rotation about an axis perpendicular to the imaging axis.
It can be shown that the ramp filter $|k_\mathrm{Dx}|$ of the backpropagation algorithm (Müller et al.\cite{Mueller15a} (equation 6.23)) becomes $|k_\mathrm{Dx}v + k_\mathrm{Dy}u|$ when the cell rotates about the axis $\mathbf{r_A}=(u,v,w)$ with $u^2+v^2+w^2=1$. In addition to this modification of the ramp filter, we introduce weights $\Delta \phi_{j}$ for all rotational positions of the cell ($j=1,2,3,\dots$) to correct for a varying speed of rotation\cite{Tam1981}.
For a set of correctly focused fields $u_{\phi_j}(\vc{r})$, measured at multiple angles $\phi_j$, the backpropagation formula reads

\begin{strip}
\begin{align}
f(\vc{r}) &= \frac{-ik_\mathrm{m}}{2\pi} 
\sum_{j=1}^{N} \! \Delta \phi_{j}  
\, D_{\mathrm{A},\phi_j} \!\!
\left\lbrace
 \text{FFT}^{-1}_\mathrm{2D}
 \left\lbrace
  \widehat{U}_{\mathrm{B},\phi_j}(k_\mathrm{Dx},k_\mathrm{Dy})
  | k_\mathrm{Dx}v + k_\mathrm{Dy}u |
  \exp[i k_\mathrm{m}(M - 1) z_{\mathrm{A},\phi_j}]
 \right\rbrace
\right\rbrace.
\label{eq:alg.backprop3d}
\end{align}
\end{strip}
Here, 
$k_\mathrm{m}= 2\pi n_\mathrm{m}/\lambda$ is the wavenumber defined by the refractive index of the medium~$n_\mathrm{m}$ and the wavelength~$\lambda$, 
$D_{\mathrm{A},\phi_j}$ is the operator for rotation about the axis \vc{r_A}, 
$\text{FFT}^{-1}_\mathrm{2D}$ is the 2D inverse Fourier transform operator,
$\widehat{U}_{\mathrm{B},\phi_j}(k_\mathrm{Dx},k_\mathrm{Dy})$ is the  Fourier transform of the background-subtracted measured field,
and $z_{\mathrm{A},\phi_j}$ parametrizes the backpropagation distance for each image within the 3D reconstruction volume.
With the rotational axis \vc{r_A} obtained from the elliptical fit, we used this formalism to reconstruct the 3D refractive index map of single cells.

\section{RESULTS} 
\begin{figure*}
\includegraphics[width=\linewidth]{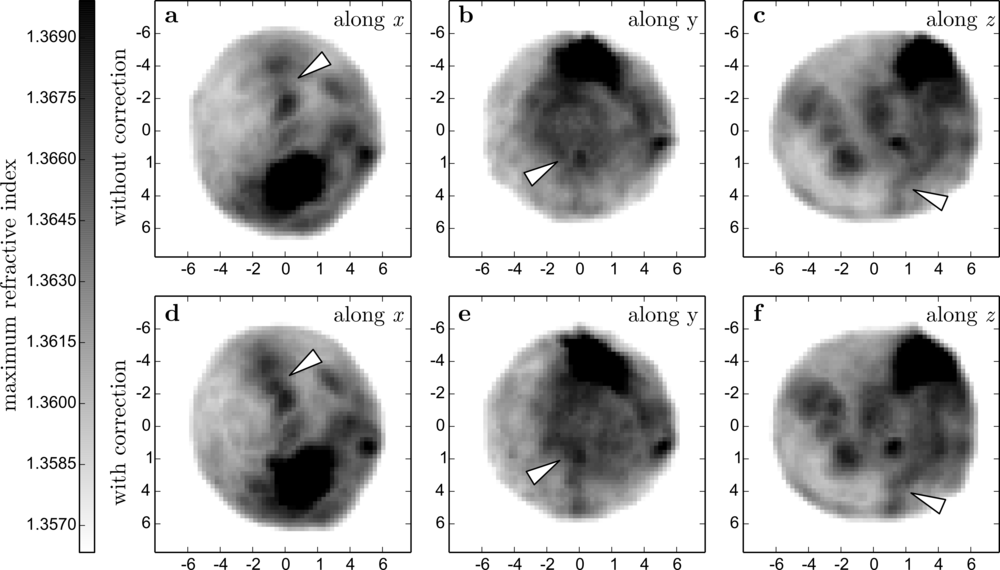}
\caption[example]
{ \label{fig:3} 
The visualization of the 3D refractive index illustrates the improvement in quality for a reconstruction \textit{without} (top row) and \textit{with} (bottom row) correction for the tilted axis of rotation. All units are in \SI{}{\um}.
The two differently reconstructed 3D refractive index distributions are visualized by maximum-projections of the refractive index along each Cartesian coordinate axis. \textbf{a,b,c)} Reconstruction of the refractive index with the assumption that the rotational axis is perpendicular to both the imaging axis and the direction of flow in the microfluidic channel. \textbf{d,e,f)} Reconstruction of the refractive index by taking into account rotation about a tilted axis. The arrows indicate areas of improved contrast.
}
\end{figure*}

Taking into account the tilted axis of rotation \vc{r_A} significantly improves the quality of the reconstructed refractive index distribution, as shown in figure~\ref{fig:3} by maximum projections along the $x$-,$y$- and $z$-axes. Here, we compare the reconstruction method with tilted-angle correction to a naive reconstruction which assumes that the cell is rotating about an axis that is parallel to the dual beam laser trap. 
The naive approach exhibits blurring as is expected for angularly misaligned projections. With the tilted-angle approach, this blurring is reduced and sub-cellular features are more visible due to improved contrast.

\section{DISCUSSION AND CONCLUSION}
The approach this paper uses to improve the refractive index reconstruction for tilted-angle rotation requires a traceable diffraction spot at the detector plane. For cells that do not exhibit such a diffraction spot, a different approach is required, for example using image cross-correlation analysis. Note that for a rotational axis that is tilted perpendicularly to the imaging plane, large tilt angles reduce the resolution of the reconstructed refractive index distribution. For example, in an extreme case where the tilt angle is 90$^\circ$ with respect to the imaging plane, the cell will rotate in the imaging plane and tomographic image acquisition will not be possible.

The rotation of the cell may be arbitrary in general, i.e. there is not necessarily a defined axis of rotation~\vc{r_A}. In those cases, a one-dimensional parametrization, covering all rotational positions of the cell, must be found to determine an equivalent of the backpropagation ramp filter $|k_\mathrm{Dx}v + k_\mathrm{Dy}u|$ discussed in this paper. Additionally, for unequally distributed angles a mechanism is required to compute the weights $\Delta \phi_{j} $ for each angular projection. The problem is equivalent to distributing a set of points on the unit sphere and dividing the entire spherical shell up into small surface area elements representing the effective area for each point. This problem can be solved by radially expanding an area element around each point on the spherical surface and defining area element edges where two areas meet.

ODT is indispensable for imaging the 3D refractive index distribution of single cells. It is a marker-free technique and thus does not suffer from issues encountered in fluorescence imaging like phototoxicity or bleaching. Furthermore, tomographic techniques produce 3D images with uniform resolution in all three spatial dimensions, which is not the case for e.g. confocal imaging with an elongated point spread function. Thus, the combination of ODT with microfluidic devices has strong implications for modern medical applications towards label-free and structure-based cell sorting.

\section*{ACKNOWLEDGMENTS}
We acknowledge helpful discussions with Ivo Sbalzarini, Pavel Tomancak, Moritz Kreysing, Martin Weigert (Max Planck Institute of Molecular Cell Biology and Genetics, Dresden, Germany), Kevin Chalut (Cavendish Laboratory, University of Cambridge, Cambridge, UK; Wellcome Trust - Medical Research Council, Cambridge Stem Cell Institute, Cambridge, UK), and Christoph Faigle (Biotechnology Center, TU Dresden, Dresden, Germany).
The HL60/S4 cells were a generous gift of D. and A. Olins (University of New England).
This project has received funding from the European Union’s Seventh Framework Programme for research, technological development and demonstration under grant agreement no 282060.

\bibliography{report}
\bibliographystyle{ieeetr}

\end{document}